\documentclass[preprint,12pt,number]{elsarticle}
\usepackage{amssymb}
\usepackage{comment}
\usepackage{multirow}
\usepackage{tabularx}
\usepackage{amsmath}
\usepackage{enumitem}
\usepackage{xcolor}
\usepackage{wasysym}
\usepackage{xurl}
\usepackage[colorlinks=true,linkcolor=blue]{hyperref}

\journal{}
\begin{document}
\begin{frontmatter}

\title{Usability Issues With Mobile Applications: \\ Insights From Practitioners and Future Research Directions}

\author[inst1]{Paweł Weichbroth*}
\affiliation[inst1]{organization={Gdansk University of Technology, Faculty of Electronics, Telecommunications and Informatics, Department of Software Engineering},
            city={Gdansk},
            country={Poland}, \newline *Corresponding Author: pawel.weichbroth@pg.edu.pl}

\begin{abstract}
This study is motivated by two key considerations: the significant benefits mobile applications offer individuals and businesses, and the limited empirical research on usability challenges. To address this gap, we conducted structured interviews with twelve experts to identify common usability issues. Our findings highlight the top five concerns related to: information architecture, user interface design, performance, interaction patterns, and aesthetics. In addition, we identify five key directions for future research: usability in AI-powered mobile applications, augmented reality (AR) and virtual reality (VR), multimodal interactions, personalized mobile ecosystems, and accessibility. Our study provides insights into emerging usability challenges and trends, contributing to both the theory and practice of mobile human-computer interaction.
\end{abstract}


\end{frontmatter}

\section{Introduction}
\label{introduction}
Mobile applications have become an increasingly important part of our business and personal lives for their convenience, accessibility and functionality \citep{islam2010mobile}. 
In business, mobile applications increase productivity by giving employees on-the-go access to critical tools and resources \citep{Wagner2023}. They also improve communication through features such as instant messaging \citep{afzal2021encrypted} and video conferencing \citep{lee2022intelligent}, and streamline processes such as project management \citep{Bali2023} or customer relationship management \citep{ahmed2014event}. 

Recent evidence also shows that mobile applications are also powerful marketing tools \citep{patsiotis2020influence}, providing a direct channel to engage with customers, deliver personalized content, and analyze user data to make informed decisions \citep{murmann2021design}. Undeniably, by taking advantage of high-speed and ubiquitous networks \citep{borcea2020mobile}, mobile applications have transformed modern business in ways that desktop digital platforms cannot \citep{turban2015mobile}.

At the individual level, mobile applications provide online services for banking \citep{alsmadi2022twenty}, shopping \citep{kim2021information}, and traveling \citep{calvignac2020can}. They also facilitate social connections through social media platforms \citep{humphreys2013mobile}, support health and well-being  goals with tracking and monitoring features \citep{pires2020research}, provide entertainment through streaming services and games \citep{falkowski2020consumption}.

Last but not least, mobile applications enable personalized education \citep{blajda2022application} and learning \citep{arain2018analysis} through online courses and freely available multimedia information resources. Certainly, mobile applications have made our daily lives easier and better in the way we communicate \citep{martin2019impact}, interact \citep{kowal2022digital} and do business \citep{ngai2007review}. It seems that mobile applications have become an integral and inseparable part of our lives \citep{bera2024exploring, Lee2025}.

Although mobile applications have been in development for more than two decades \citep{Pittman2020}, they continue to face numerous usability issues that can negatively impact users' intention to use \citep{coursaris2012impact}. In addition, dissatisfied users often seek alternative solutions, eventually leading to the abandonment of such applications. This highlights the timely need for research on mobile usability as the development of effective, efficient, and engaging mobile applications still remains a relevant endeavor \citep{coursaris2011meta}.

While numerous studies have documented user feedback on mobile usability issues, to our knowledge, relatively little research has focused on the perceptions of stakeholders involved in the design and development process. This study aims to address this gap by examining the topic from the perspective of practitioners in the mobile application development industry. Given its nature, a qualitative study is the obvious choice, with all the implications that entails.

The rest of this paper is structured as follows. 
Section~\ref{sec:background} provides the background and motivation, necessary to understand the research topic as well as briefly discuss its rationale. 
Section~\ref{sec:methodology} presents the applied research methodology, including detailed specifications of the settings, as well as the data collection and analysis approach. 
Section~\ref{sec:results} discusses the results, focusing on the extracted issues and introducing broader related categories.
Section~\ref{sec:discussion} reflects on the limitations of the study and its specific contributions to theory and practice. 
Section~\ref{sec:future-research-directions} outlines future research directions that may be considered and pursued by other researchers and practitioners.

\section{Background \& Motivation}
\label{sec:background}
The global mobile market was valued at approximately \$252.89 billion in 2023 and is expected to grow at a compound annual growth rate of 14.3\% to reach an estimated \$626.39 billion by 2030 \citep{GVR2025}. This growth is being driven by factors such as the widespread adoption of smartphones, increased Internet penetration, and the availability of numerous online services \citep{Natanson2021}. 

The shift from desktop computers, typically equipped with a large screen, full-size keyboard, and external mouse, to average-sized smartphones with screens of about six inches has introduced a new set of conditions that are not always perceived as comfortable or efficient \citep{gafni2009usability}. For software vendors offering mobile applications \citep{Griffiths2015}, usability has once again taken center stage, drawing considerable attention from the research community as well \citep{nayebi2012state}. 

The effort devoted to mobile usability research has two streams. While the first focuses on theory, the second is devoted to practice. In particular, the former aims to develop frameworks, models, and principles, while the latter focuses on testing and evaluating the usability of real-world mobile applications. In a natural and seamless manner, theory and practice intertwine, each complementing and strengthening the other. 

By design, usability testing is intended to obtain feedback from users, evaluate its effectiveness and efficiency, and identify and classify the issues encountered and reported. The rationale behind testing involves many factors.
Usability issues not only affect user satisfaction \citep{ferreira2020impact}, but also have a tangible impact on business results \citep{konradt2012role}. For example, mobile applications that suffer from usability issues can result in decreased conversion rates \citep{daoud2023mobile}, lower customer retention \citep{wahab2011influence}, and diminished brand reputation \citep{casalo2007role}.

On the other hand, addressing these issues can lead to improved metrics such as higher conversion rates, increased user retention, and positive word-of-mouth \citep{rajaobelina2021relationship}. 
Note that in a broader sense, investing in usability testing is said to improve the overall performance and profitability of digital products \citep{byun2020evaluation, gatsou2013exploring}. Basically, usability testing is an important part of software quality assurance \citep{dhillon2022applied}.

Usability as such is not an abstract concept. Recent research \citep{weichbroth2020usability} shows that to specify usability in a mobile context, the most commonly used standard is ISO 9241--11, which defines its scope as follows: "extent to which a product can be used by certain users to achieve certain goals with effectiveness, efficiency, and satisfaction in a certain context of use" \citep{ISO9242-11}. In a narrow sense, there are three different qualities that are the subject of evaluation.

In fact, over the years, more than 40 different usability attributes have been "invented" and tested in different settings and circumstances \citep{weichbroth2020usability}, in order to evaluate perceived usability in a more comprehensive and effective way. However, the vast majority of studies have focused on gathering feedback from users, with only a few involving practitioners working in the mobile application development business. This paper attempts to fill this gap by sharing the results of a survey by highlighting their perspectives and insights on the usability issues. 

\section{Methodology}
\label{sec:methodology}

\subsection{Settings}
By definition, an interview is a qualitative research method that relies on asking questions to gather primary data \citep{alshenqeeti2014interviewing}.
Since our study has a specific and precise objective, we used a form of structured expert interview without the need for a follow-up discussion and evaluation. This type of data collection method is particularly useful when investigating a specific experience or phenomenon and can facilitate the analysis of the results, since the questions are predetermined. 
It should also be highlighted that expert interviews are now widely used as a qualitative research tool in usability research \citep{hirai2007evaluation, van2017barriers}, and more recently in the mobile context \citep{jeong2020gui, wich2015enhanced}. 

For us, an expert is a person who is responsible for a concept, an implementation, or a problem-solving capability, as someone who has relevant factual, generic, or specific knowledge about processes, group behaviors, strategic decisions related to mobile software products or services. 
In this view, an expert is a person with technical, process, and interpretive knowledge. In the context of the current study, an expert is a person with proven professional experience, demonstrated by at least one project in the area of mobile application development. Specifically, an expert must have at least three years of work experience in a role such as UI or UX designer, graphic or product designer, front-end developer, or any other relevant role considering the scope of the position's responsibilities. 

\subsection{Data Collection}
We reached the group of experts through social media networking tools. The initial screening of the available information from each profile allowed us to select the preliminary group of twenty candidates who met the inclusion criteria. We then made initial contact by sending an email or message with a research agenda and a polite request for participation, emphasizing that the interview should not exceed 15 minutes and that all personal information would be disclosed. 

In the end, twelve experts (five woman, seven men) agreed to participate. A summary of the experts included in the study is shown in Table \ref{tab:sample-desc}.

\begin{table*}[h]
\caption{Summary of demographic information and professional background of the experts.}
\label{tab:sample-desc}
\centering
\footnotesize	
\begin{tabular}{|l|l|l|l|l|l|l|}
\hline
\textbf{Id} & \textbf{Gender}  & \textbf{Age} & \textbf{Education} & \textbf{Current Occupation}  & \textbf{\# Experience} & \textbf{\# Projects} \\ \hline
EX01      & Woman & 33  & Higher    & UX Designer                & 3          & 3        \\ \hline
EX02      & Man   & 26  & Secondary & Front-End Developer        & 4          & 3        \\ \hline
EX03      & Man   & 39  & Higher    & Front-End Developer        & 16         & 4        \\ \hline
EX04      & Man   & 32  & Higher    & Software Developer         & 10         & 4        \\ \hline
EX05      & Woman & 45  & Higher    & UX Designer                & 20         & 19       \\ \hline
EX06      & Man   & 25  & Higher    & Senior Data Engineer       & 6          & 3        \\ \hline
EX07      & Woman & 56  & Higher    & Software Developer         & 20         & 15       \\ \hline
EX08      & Woman & 30  & Higher    & UX Designer                & 8          & 9        \\ \hline
EX09      & Man   & 42  & Higher    & UX Designer                & 17         & 25       \\ \hline
EX10      & Woman & 46  & Higher    & Front-End Developer        & 20         & 15       \\ \hline
EX11      & Man   & 29  & Higher    & IT Team Leader             & 3          & 2        \\ \hline
EX12      & Man   & 31  & Secondary & UX Designer                & 7          & 3        \\ \hline
\end{tabular}
\end{table*}

The mean age of the study participants was 36.17 years. The vast majority (10 out of 12) claimed to have a higher education, while two (both male) had a secondary education. All experts met the requirements in terms of position, work history ($\geq 3$ years), and work experience ($\geq 1$ project). 

For the purpose of the survey, we prepared an electronic form consisting of three parts. The first, the introduction, contains the guiding theme of the survey and a brief introduction to the survey. The second part presents a research question formulated as in the following way: What are the most common usability issues in mobile applications?
  
The third part asks for the expert's demographics, including age, education, current occupation, years of experience, and a request to identify and describe IT projects the expert has been involved in over the past three years. 

From February to December 2024, we sent a message (email) to each of the experts asking them to respond independently within one week, with no word limit for each question. It should be noted that the research question was designed to explore the research phenomenon in an open-minded way and to understand how experts, in the context of their work environment, formulate and perceive usability issues. 

In addition, the experts were also aware that they would be given the opportunity to end the interview with any comments or questions. We therefore adopt an interpretive stance, seeking to understand the approaches and activities undertaken, as well as the reported experiences and perceptions at the individual level. In our opinion, this gave us a broad overview and deep insights into the premises and the emerging usability issues.

The expert feedback was collected in Polish. They were then translated into American English. Two online translators were used: Google Translate (https://translate.google.com/) and DeepL (https://www.deepl.com/pl/translator). Both are available to the public and free of charge. Each statement was translated twice, independently by both translators. The translations were then compared to evaluate the terminology used. The quality of the translations was satisfactory, and only two required minor changes.

In quantitative terms, Table \ref{tab:data-quant} summarizes the collected responses of all experts, along with an estimate of the number of words delivered.

\begin{table}[h]
\caption{Data quantification of expert responses in terms of gender, number of words, and their percentage of a collected data set.}
\label{tab:data-quant}
\footnotesize
\centering
\begin{tabular}{|l|l|l|l|}
\hline
\textbf{ID}    & \textbf{Gender}   & \textbf{\#Words} & \textbf{Share}    \\ \hline
EX01 & Woman & 84      & 7.08\%   \\ \hline
EX02 & Man   & 172     & 14.50\%  \\ \hline
EX03 & Man   & 137     & 11.55\%  \\ \hline
EX04 & Man   & 11      & 0.93\%   \\ \hline
EX05 & Woman & 106     & 8.94\%   \\ \hline
EX06 & Man   & 93      & 7.84\%   \\ \hline
EX07 & Woman & 11      & 0.93\%   \\ \hline
EX08 & Woman & 172     & 14.50\%  \\ \hline
EX09 & Man   & 98      & 8.26\%   \\ \hline
EX10 & Woman & 131     & 11.05\%  \\ \hline
EX11 & Man   & 125     & 10.54\%  \\ \hline
EX12 & Man   & 46      & 3.88\%   \\ \hline
\end{tabular}
\end{table}

A total of 1186 words were collected from twelve experts, with an average of 98.83 words per individual response. The proportion of women (42.50\%) is lower than that of men (57.50\%), whereas the average is 3.37\% higher.
The most productive was expert eight (\#172), and the least productive was expert four (\#11). The quantitative evaluation was also confirmed by the qualitative analysis which shows that the longer the statement, the higher its value. 

\subsection{Data analysis}
The collected data were subjected to further analysis. Careful reading was done to code the data. By definition, to code means "to assign a truncated, symbolic meaning to each data point for purposes of qualitative analysis" \citep{leavy2014oxford}. 
Now, considering the nature of the formulated research question, in vivo coding was applied which "refers to a code based on the actual language used by the participant" \citep{leavy2014oxford}. 
In this regard, we followed the guidelines elaborated by Theron \citep{theron2015coding}, while the systematic approach was adopted from Arije et al. \citep{arije2021synthesizing}. 
Therefore, in our study the system for inductive thematic content analysis consisted of three stages:
\begin{enumerate}
    \item Coding, extracting exact words as codes to capture the lived perspectives and experiences of the participants.
    \item Abstracting, naming the extracted codes to communicate their essential meaning through valid terminology.
    \item Synthesizing, grouping and integrating findings to draw general insights.
\end{enumerate}

In the first stage, data coding involved three steps, namely: (\textit{a}) extract codes from the text which means in vivo coding to separate text strings with consistent meaning without changing the original spelling; (\textit{b}) analysis and evaluation of the relevance of the codes: separation of individuals, and (\textit{c}) remove irrelevant codes from the data volume. In this extent, Table \ref{tab:data-coding-RQ1} shows the particular results in quantitative terms.

\begin{table}[h]
\caption{Results of data coding and detailed analysis.}
\label{tab:data-coding-RQ1}
\footnotesize
\centering
\begin{tabular}{|l|l|l|l|}
\hline
\textbf{ID}    & \textbf{\#Extracted} & \textbf{\#Removed} & \textbf{\#Remained} \\ \hline
EX01 & 25          & 3         & 22         \\ \hline
EX02 & 25          & 8         & 17         \\ \hline
EX03 & 22          & 20        & 2          \\ \hline
EX04 & 4           & 2         & 2          \\ \hline
EX05 & 14          & 6         & 8          \\ \hline
EX06 & 15          & 8         & 7          \\ \hline
EX07 & 4           & 0         & 4          \\ \hline
EX08 & 28          & 11        & 17         \\ \hline
EX09 & 15          & 6         & 9          \\ \hline
EX10 & 20          & 16        & 4          \\ \hline
EX11 & 17          & 11        & 6          \\ \hline
EX12 & 10          & 2         & 8          \\ \hline
\textbf{Sum}   & \textbf{199 }        & \textbf{93}        & \textbf{106}        \\ \hline
\end{tabular}
\end{table}

As can be noted, a total of 199 codes were extracted from the first data set, including 91 (45.73\%) from female and 108 (54.27\%) from male participants. The largest number of 28 codes was extracted from the answer given by EX08, and 25 from both EX01 and EX02. In contrast, the smallest number of four codes was generated from expert responses EX04 and EX07. Each subset was then subjected to detailed analysis and evaluation based on the criteria of merit, relevance and validity. 

In total, 93 (46.73\%) codes were removed from the original dataset, including 36 (38.71\%) from the female group and 57 (61.29\%) from the male group. Note that the relative dropout was lower for the former group (39.56\%) compared to the latter (52.78\%). Thus, it seems that a female group delivered more concise and substantial content than a male group.
On an individual level, the largest number of codes removed was 20 (90.91\%) from EX03, 16 (80\%) from EX10, and 11 from both EX08 (39,29\%) and EX11 (64,71\%). 

For example, one WCAG code was removed because it stands for Web Content Accessibility Guidelines, which is an international standard not strictly related to mobile usability issues. Other codes classified as irrelevant included: "acceptable level", "afraid to use application", "missing applications", "mobile market standards", and "Sudoku app example", to name a few.
Finally, 106 (53.27\%) codes remained for further investigation.

In the second stage, we distinguished codes that form an unambiguous term consisting of one word (e.g. "navigation", "performance"), or two or more words (e.g. "color scheme"). In addition, we manually abbreviate codes to a recognized general term by removing adjectives nouns that contain obvious or redundant information from the point of view of analysis. For example, while smartphone is typically used by a person, "security of personal data" has been changed to Security. Other changes included, for instance "application performance" and "quality performance" to Performance. A total of 15 unique terms were identified, covering 23 codes.

In addition, we removed sentiment keywords, reflecting subjective judgment with respect to particular artifact. For instance, "small buttons" and "unclear buttons" to "buttons", or "illegible fonts" to "fonts". Therefore, 15 unique terms were identified, including 39 codes. 

In third stage, we performed semantic analysis, based on the existing terms to avoid conceptual redundancy (when it was possible), the remaining codes have been interpreted with the aim of grouping together different inflected forms of the same term. For instance, the code "frequent application crashes" and "not responding in expected time" have been both changed to Performance, or "bottom bar ads" to Advertising, or "too many features", "difficult to find information" and "no logical structure" were all grouped under Information Architecture. In such manner, 44 codes were grouped under 8 recognized terms.

\begin{figure}[ht]
    \centering
    \includegraphics[width=0.85\linewidth]{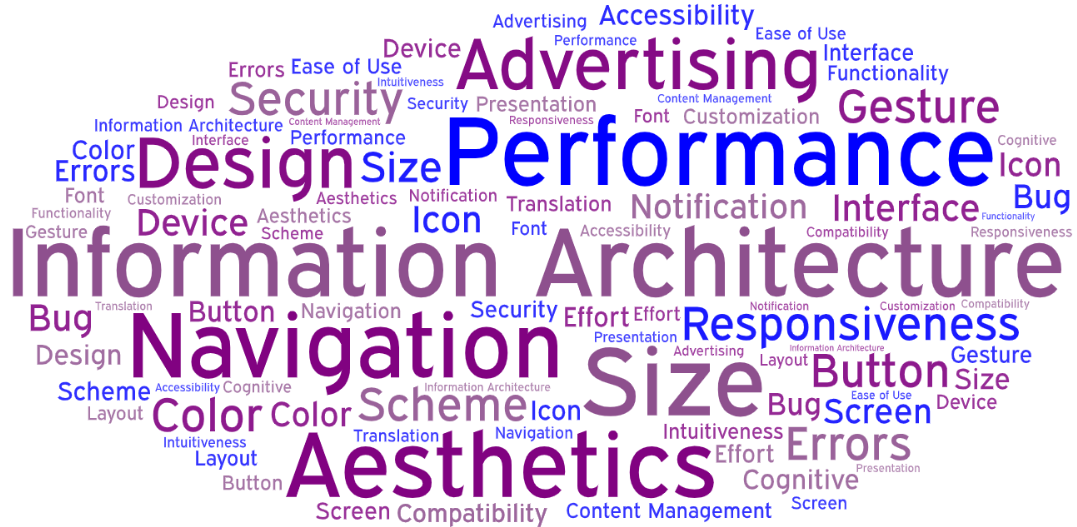}
    \caption{A cloud of 30 terms describing mobile usability issues.}
    \label{fig:RQ1-terms}
\end{figure}

In summary, a set of 30 unique terms were identified (see Figure~\ref{fig:RQ1-terms}) that cover all 106 extracted codes. This set is further considered and discussed in the next section, as a basis for formulating the answer to the first research question.

\section{Results}
\label{sec:results}
Table \ref{tab:terms-identified} shows the set of 30 terms as a result of the qualitative analysis. Note that experts used these terms, in conjunction with other words, in different configurations and contexts. As can be seen, the diversity of terms suggests that mobile issues span multiple dimensions, going well beyond the boundaries of usability and into the realm of user experience.

\begin{table*}[h]
\caption{Extracted terms from expert responses with number of codes assigned}
\label{tab:terms-identified}
\footnotesize
\centering
\begin{tabular}{|l|l|l|l|l|l|}
\hline
\textbf{No} & \textbf{Term}   & \textbf{\#Codes} & \textbf{No} & \textbf{Term}  & \textbf{\#Codes} \\ \hline
1  & Information Architecture & 24      & 16 & Icon               & 2       \\ \hline
2  & Performance              & 17      & 17 & Notification       & 2       \\ \hline
3  & Navigation               & 6       & 18 & Screen Size        & 2       \\ \hline
4  & Aesthetics               & 5       & 19 & Accessibility      & 1       \\ \hline
5  & Advertising              & 4       & 20 & Cognitive Effort   & 1       \\ \hline
6  & Design                   & 4       & 21 & Compatibility      & 1       \\ \hline
7  & Responsiveness           & 4       & 22 & Content Management & 1       \\ \hline
8  & Security                 & 4       & 23 & Ease of Use        & 1       \\ \hline
9  & Button                   & 3       & 24 & Customization      & 1       \\ \hline
10 & Color Scheme             & 3       & 25 & Font               & 1       \\ \hline
11 & Errors                   & 3       & 26 & Functionality      & 1       \\ \hline
12 & Gesture                  & 3       & 27 & Intuitiveness      & 1       \\ \hline
13 & Interface                & 4       & 28 & Layout             & 1       \\ \hline
14 & Bug                      & 2       & 29 & Presentation       & 1       \\ \hline
15 & Device Size              & 2       & 30 & Translation        & 1       \\ \hline
\end{tabular}
\end{table*}

Figure \ref{fig:RQ1-terms} shows the set of 30 terms as a result of the qualitative analysis. Note that experts used these terms in different configurations and contexts. As can be seen, the diversity of keywords suggests that mobile issues span multiple dimensions, going well beyond the boundaries of usability and into the realm of user experience.

Certainly, they cannot be dissected separately, given the intangible nature of the mobile user experience. On the other hand, it can be concluded that the multifaceted its nature  implies the need to address a wide range of interrelated factors that influence the interaction and, consequently, a variety of issues that can arise. 

Having said that, in order to simplify and generalize the findings, we decided to arrange identified terms into categories. Here, by a category is a group of terms that share some commonality. Thus, a category is a broader concept, while comparing to a term. However, given the comprehensive nature of some of the terms already discussed, only two new categories were used, viz: Hardware and Interaction Patterns. 
The term grouping concerned the following five categories:
\begin{itemize}
    \item Information Architecture: Accessibility; Cognitive Effort; Intuitiveness; Content Management; Translation.
    \item Interface: Design; Button; Color Scheme; Icon; Customization; Font; Layout; Presentation; Ease of Use; Functionality.
    \item Interaction Patterns: Gesture; Notification; Navigation.
    \item Errors: Bug.
    \item Hardware: Device Size; Compatibility; Screen Size.
\end{itemize}

\begin{figure}[h]
    \centering
    \includegraphics[width=0.95\linewidth]{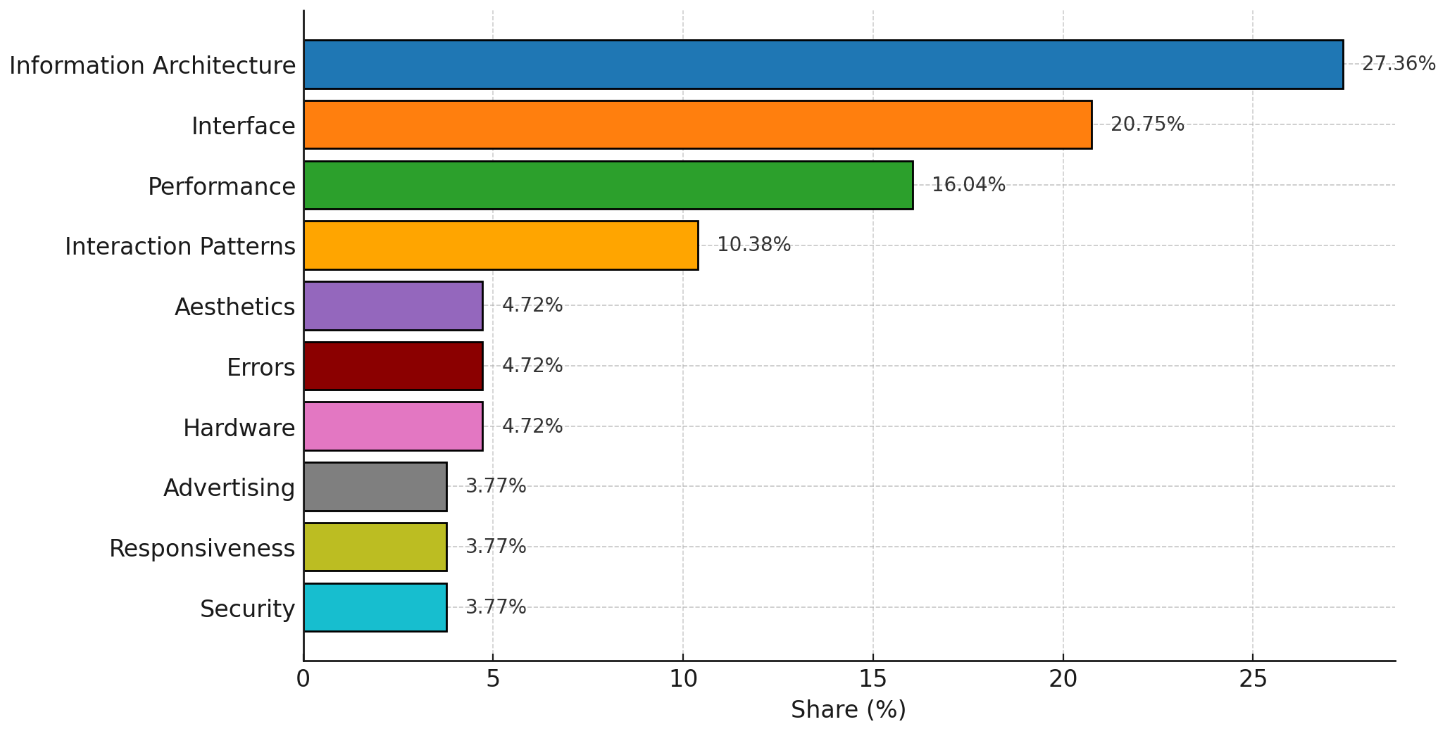}
    \caption{Distribution of categories related to mobile application usability issues.}
    \label{fig:categories-share}
\end{figure}

Table~\ref{tab:RQ1-resutls} lists the set of ten general categories related to mobile usability issues, along with the number of associated codes shows the strength of support across the group of experts who participated in our study. In addition, 41 specific mobile usability issues were extracted from the data collected. Each category, including associated usability issues, is briefly outlined below (Figure~\ref{fig:categories-share}).

\begin{table}[h]
\caption{Categories of mobile usability issues, along with the number codes assigned as the number of specific issues reported by the expert panel.}
\label{tab:RQ1-resutls}
\footnotesize
\centering
\begin{tabular}{|l|l|l|l|l|}
\hline
\textbf{No} & \textbf{Category}    & \textbf{\#Codes} & \textbf{Share}    & \textbf{\#Issues} \\ \hline
1  & Information Architecture & 29      & 27.36\%  & 10        \\ \hline
2  & User Interface           & 22      & 20.75\%  & 9        \\ \hline
3  & Performance              & 17      & 16.04\%  & 6        \\ \hline
4  & Interaction Patterns     & 11      & 10.38\%  & 3        \\ \hline
5  & Aesthetics               & 5       & 4.72\%   & 4        \\ \hline
6  & Errors                   & 5       & 4.72\%   & 2        \\ \hline
7  & Hardware                 & 5       & 4.72\%   & 3        \\ \hline
8  & Advertising              & 4       & 3.77\%   & 4        \\ \hline
9  & Responsiveness           & 4       & 3.77\%   & 2        \\ \hline
10 & Security                 & 4       & 3.77\%   & 0        \\ \hline
\end{tabular}
\end{table}

As can be seen, Information Architecture (IA) (27.36\%) takes the first place among all other categories. In general, IA refers to the organization of information, focusing on content structure and presentation. However, in the context of mobile applications, IA also refers to objects that reflect an application's features. 
While information architecture was reported twice directly as a problem, the other most common usability issues that fall into this category are: content overload (\#5), abundance of content (\#5), application complexity (\#4), difficulty finding information (\#3), lack of clear instructions or feedback (\#3), lack of logical structure (\#3), layout inconsistency (\#1), and difficulty finding features (\#1). Furthermore, content management (\#1) and translation (\#1) were also pointed out as obstacles to users.

User Interface (UI), considered in terms of the graphical user interface (GUI), is the second (20.75\%) largest group of usability issues for mobile applications. In practice, the GUI can be seen as the materialization of an underlying information architecture pattern. While the GUI enables user interaction with an application, the problems arise from both the interaction itself and the components used. 
Not surprisingly, the GUI was directly mentioned as a usability problem four times. The other more specific issues are relate to: design (\#4), buttons (\#3), color scheme (\#3), icons (\#2), fonts (\#1), regarding their size and location. Note that others concern: presentation (\#1), along with customization (\#1), functionality (\#1), and ease of use (\#1).

Performance is the third (16.04\%) most discussed category by experts. Note that in the realm of usability, performance is typically considered in terms of application efficiency, primarily speed and operational stability, as well as measured by the consumption of available hardware resources (battery, CPU, memory) and network bandwidth. 
Here, the most common issues concern: response time (\#4), speed of loading content (\#2), crashes (\#1), 
sudden battery loss (\#1), error handling (\#1), and inability to adapt to new environment conditions (\#1). 
Notably, performance as such was also raised as a general issue seven times, mostly due to poor network availability. 

Interaction Patterns, the fourth (10.38\%) category in the list of mobile usability issues, refers to a solution for how users interact with an application's interface to accomplish specific tasks or goals. Currently, two methods are in common use: gestures and spoken commands. The former refers to a physical action performed by the user on a touch-sensitive screen, including pinch, tap, scroll, swipe, rotate, and zoom.
In this view, the reported issues are related to navigation (\#6) and gestures (\#3). Considering the latter method, none of the obstacles were raised. In addition, a push notification mechanism (\#2), responsible for sending a clickable pop-up messages directly to the user's smartphone, poses significant obstacles in some applications. 

The next three categories on the list with less than five percent (4.72\%) are (in alphabetical order): Aesthetics, Errors, Hardware.
Aesthetics refers to the visual qualities that influence the user's emotional perception of the mobile application. In this regard, usability is compromised by lack of simplicity (\#2) or harmony in layout (\#1), excessive flashiness (\#1), and visual distractions from content (\#1).

An error is any software bug, flaw, failure, or fault in a mobile application that causes it to produce an incorrect or unexpected result. An application error can affect a user's performance in one of two ways: the user may or may not be able to continue and complete a task. In our study, errors (\#3) and bugs (\#2) were generally highlighted as barriers and issues that affect mobile usability.

Hardware issues and barriers include malfunctions with the physical components of the mobile device, typically related to the display (\#2), battery (\#2), and compatibility (\#1). Thus, note that these factors can be considered external, as they are neither the application attributes nor the user perceptions, given the scope of mobile usability. 

Advertising, Responsiveness, and Security are the last three categories, each grouping the smallest number (3.77\%) of codes extracted. 

By definition, advertising is a form of communication used to promote products or services, and typically its message evokes a neutral or positive emotional response from its recipients. However, in the context of this study, unsolicited and embedded ads are claimed to negatively affect the user perception and to be a usability issue due to their clutter (\#1), disruptive behavior (\#1), or intrusive placement (\#1). On the other hand, bottom bar ads (\#1) are also highlighted as a mobile usability issue.

A responsive app can be viewed and used on a variety of devices, from desktops to tablets to smartphones. In fact, responsive apps are web apps that launch through web browsers. Keeping in mind that one size doesn't fit all, responsiveness was brought up twice (\#2). More specifically, experts also pointed out issues related to layout (\#1) and screen size (\#1) optimization.

According to the experts, security is no longer just a technical imperative, but a critical privacy requirement as mobile applications are increasingly relied upon for both personal and business financial activities. From this perspective, insecure mobile solutions that are vulnerable to data breaches or unauthorized access were labeled as hazardous (\#4) and eventually pointed out as unacceptable to release from a software vendor perspective.

\section{Discussion}
\label{sec:discussion}
Obviously, experts involved in different mobile application development projects have encountered different kinds of issues, obstacles, and limitations. Although there is no consensus among the experts, it was possible to extract a common part through a detailed contextual analysis of the sentences in a narrow view based on in vivo data coding. 
However, this approach comes with inherent limitations. 

The first concern is the use of interviews as a research method to collect data, which consequently imposes several limitations that should be considered when interpreting and evaluating the results. In this view, a major limitation is the potential for subjectivity and bias, as both interviewers and respondents can influence the data through confirmation bias or the use of certain technologies in the design or implementation of an application. On the other hand, participants may avoid highlighting certain issues for which they have partial or full responsibility, which may affect the authenticity and reliability of their responses.

The second concern is the limited generalizability of the interview results. Since our study relies on a small, non-random sample, the findings cannot be considered representative of the broader population. Nonetheless, to the best of our knowledge, the reported issues pertain to a few mobile solutions that have reached a global audience, including non-native Polish users. Therefore, the findings discussed are not context specific and can still be considered applicable to other groups or settings.

The third concern is the interpretation of interview data, which presents its own set of challenges. The richness of qualitative responses can make it difficult to extract clear and concise conclusions, and researchers may inadvertently impose their own perspectives or biases during analysis. It is important to acknowledge the use of automated translation, which may not always fully capture the intended meaning due to a lack of contextual understanding. However, we found that the outcomes of these AI-powered tools were highly accurate and precise, effectively preserving both terminology and context.

Last but not least, reaching experts in the desired field is challenging, and convincing them to share their experiences is even more difficult. This issue arises not only from their limited availability but also from restrictions on formal disclosure of sensitive information that may emerge during in-person interactions. As a result, participants' personal information, including the names of their employers, projects, mobile solutions, and technologies used, cannot be shared with third parties or discussed within the study. 

Despite the above limitations, we remain confident that the results obtained have a high degree of relevance and validity.
In this sense, our study contributes to the theory by identifying and specifying the set of usability issues related to mobile applications based on evidence-based primary data. In addition, these issues were organized into 10 categories, thus introducing a novel classification. Considering its design, a theory of mobile human-computer interaction is updated through by factual data that can be further used in developing new concepts and frameworks. While their relevance is confirmed not only by ongoing technological advances, but more importantly by the constantly evolving needs and requirements of mobile application users, their validity is reflected in their strong compliance with the state of the art knowledge.

Moreover, our study provides significant practical value to software vendors by highlighting specific issues that have the potential to make a difference in the user experience of mobile applications offered in the marketplace. From this perspective, the presented catalog informs a wide range of interested parties, from UI/UX product designers and software engineers to product owners and project managers, in a straightforward and understandable manner, as it relies on the expertise of their counterparts.  Ultimately, this research provides the guidelines for usability testing. Moreover, it can also provide an empirical foundation to refine existing as well as develop new quality control checklists.

Nonetheless, more research is still needed as the marketplace for mobile applications continues to grow. In December 2024 alone, approximately 41k and 38k mobile apps were released on the Google Play Store \citep{Statista-Google-2024-XII} and the Apple App Store \citep{Statista-Apple-2024-XII}, respectively. On the other hand, market saturation with a wide range of substitutes leads to constantly rising quality expectations from users. This not only presents immense challenges for software vendors, but also attracts the attention of the research community interested in timely and impactful research opportunities. 

In summary, we believe that the empirical evidence presented in this paper offers valuable insights for both theory and practice. By exploring the experiences and challenges shared by each expert, this study aims to identify common patterns and insights that bridge the gap between theoretical principles and practical implementations.

\section{Future Research Directions}
\label{sec:future-research-directions}
On the basis of both the interviews and the most recent research from the past three years, five significant and ongoing lines of research have emerged in the foreground:
\begin{itemize}
    \item Usability in AI-powered mobile applications \citep{deniz2023quality, nama2023ai, namoun2024predicting}. As AI-driven mobile apps (e.g., chatbots, contextual recommendations, personalized agents) become more prevalent, research efforts can be undertaken to investigate new, more intelligent forms of interaction.

    \item Usability in Augmented Reality (AR) \citep{sung2021effects, yavuz2021augmented, criollo2021towards} and Virtual Reality (VR) \citep{kim2021applications, chan2024using, omran2024virtual}. AR and VR bridge the digital and physical worlds by integrating digital visual and audio content into the user's real-world environment in real time. Undeniably, the emergence of these technologies opens up new frontiers for usability research.

    \item Usability in Multimodal Interactions \citep{oviatt2022multimodal, azofeifa2022systematic, dritsas2025multimodal}. The shift to multimodal interfaces that integrate speech, gesture, touch, and eye tracking is no longer a theoretical notion \citep{Carter2024}. More research is needed to evaluate how these modalities can be effectively combined to provide natural and intuitive ways to interact seamlessly with a user.

    \item Usability in Personal Mobile Ecosystems \citep{bender2021impact, cohen2021urban, eom2023effects}. This concept goes beyond interacting with a single mobile device by leveraging all available, paired, and connected devices \citep{mannonen2013approach}, including smartphones, tablets, wearables, smart home devices, smart car systems, and even smart audio devices. While mobile applications consume resources from more than a single device, usability testing requires a major reframing. 
    
    \item Usability vs Accessibility \citep{senjam2021smartphones, alajarmeh2022extent, zaina2022preventing}. While built-in accessibility features have become an integral part of mobile applications, driven by advanced technologies from Apple \citep{Carter2024b}, Huawei \citep{huawei2025} or IBM \citep{Takagi2021}, the current methods and frameworks for usability testing and evaluation require a major revision.
\end{itemize}

Having said that, we can conclude that the mobile usability landscape requires a comprehensive review, including both theoretical foundations and practical methods and tools. On the other hand, while the adoption of innovative technologies offers considerable benefits, it is equally important to recognize the associated risks and threats. In summary, we are confident that with careful planning, implementation, and testing, while prioritizing user security, privacy, and usability, any technology can be effectively leveraged to deliver meaningful benefits.

\section*{Author Contributions}
Paweł Weichbroth: Conceptualization, Methodology, Validation, Formal analysis; Investigation; Resources; Data Curation; Writing (Original Draft); Writing (Review \& Editing); Software; Visualization; Supervision; Project administration; Funding acquisition.

\section*{Data availability}
Data available on reasonable request.

\section*{Funding}
The research was supported in part by project “Cloud Artificial Intelligence Service Engineering (CAISE) platform to create universal and smart services for various application areas”, No. KPOD.05.10-IW.10-0005/24, as part of the European IPCEI-CIS program, financed by NRRP (National Recovery and Resilience Plan) funds.


\section*{Declaration of Interest Statement}
The author declares that he has no known competing financial interests or personal relationships that could have appeared to influence the work reported in this article.

\bibliographystyle{authordate1}

\end{document}